\documentstyle[12pt]{article}
\def\beq{\begin{equation}}
\def\eeq{\end{equation}}
\def\bea{\begin{eqnarray}}
\def\eea{\end{eqnarray}}
\def\bq{\begin{quote}}
\def\eq{\end{quote}}
 
\begin{document}
\pagestyle{empty}
\begin{flushright}
\end{flushright}
\vspace*{5mm}
\begin{center}
{\bf TWISTED LOCAL SYSTEMS SOLVE THE (HOLOGRAPHIC) LOOP EQUATION OF 
LARGE-$N$ $QCD_4$ }
\\  
\vspace*{1cm} 
{\bf M. Bochicchio} \\
\vspace*{0.5cm}
INFN Sezione di Roma \\
Dipartimento di Fisica, Universita' di Roma `La Sapienza' \\
Piazzale Aldo Moro 2 , 00185 Roma  \\ 
e-mail: marco.bochicchio@roma1.infn.it \\
\vspace*{2cm}  
{\bf ABSTRACT  } \\
\end{center}
\vspace*{5mm}
\noindent
 We construct a holographic map from the loop equation of large-$N$ $QCD$
 in $d=2$ and $d=4$, for planar self-avoiding loops, to the critical equation
 of an equivalent effective action. The holographic map is based on two
 ingredients: an already proposed holographic form of the loop equation, such that
 the quantum contribution is reduced to the evaluation of a regularized residue;
 a new conformal map from the region encircled by the based loop to a cuspidal
 fundamental domain in the upper half-plane, such that the regularized residue
 vanishes at the cusp which is the image of the base point of the loop.
 The critical equation of the holographic effective action determines a unitary
 Abelian local system in $d=2$ and a non-Abelian twisted local system in $d=4$.
 As a check in the $d=2$ theory, we study the distribution of eigenvalues of the
 Wilson loop implied by the critical equation.
 As a check in the $d=4$ theory, we study the first coefficient of the beta function
 implied by the holographic loop equation and, as a preliminary step, that part of
 the second coefficient which arises from the rescaling anomaly, in passing from the
 Wilsonian to the canonically normalised (holographic) effective action.
\vspace*{1cm}
\begin{flushleft}
\end{flushleft}
\phantom{ }
\vfill
\eject
\setcounter{page}{1}
\pagestyle{plain}

\section{Introduction}

This paper grew out of the attempt to obtain directly from the loop
equation \cite{MM,MM1} for self-avoiding loops the celebrated formula for the
distribution of the eigenvalues of a Wilson loop of area $A$ in the weak coupling phase,
in the large-$N$ limit of two dimensional $QCD$ on a sphere,
for $g^2 A$ small with respect to $g^2 A_{sphere}$:
\bea
W(A)= \int \prod d\theta_i \prod_{i \ne j}|\theta_i-\theta_j|
\exp(-\frac{N}{2g^2A} \sum_i \theta_i^2) \times \nonumber \\
\times \sum_i N^{-1} \cos \theta_i
\eea
An essentially equivalent formula was first obtained from the Wilsonian
lattice action using functional
techniques \cite{GW} and afterward Eq.(1) was derived from the
"heat kernel action" \cite{D2,KA} or from the semi-circle law for the eigenvalues
of free random variables \cite{V}, which are employed in the complete operatorial
solution of the loop equation in $d=2$ for arbitrary self-intersections \cite{G}. 
We present here a derivation based on the loop equation restricted to 
self-avoiding loops but that extends to the (planar) four dimensional
case. Restricting to (planar) self-avoiding loops has some virtue, since the general
solution for arbitrary self-intersection seems to be out of the reach of our
methods in $d=4$ and in $d=2$ as well.
In this respect, already long ago, the loop equation in $d=2$ and in $d=4$
has been written in terms of the distribution of the eigenvalues for self-avoiding
loops \cite{Duh} \footnote{We thank the referee for pointing out ref. \cite{Duh} to our
attention.}. In Eq.(1) and in this paper we have ignored
the contribution of the exterior of the loop, assuming that the internal region is much
smaller than its complement. In addition we have not taken into account the periodicity
of the eigenvalues. We believe that the techniques of this paper can be extended
to include also the more general case. However for simplicity we will consider it
elsewhere.
In our derivation there are two basic ingredients. The first one is a holographic
form of the loop equation for planar loops in $d=2$ and $d=4$.
This holographic loop equation was derived in \cite{MB}. 
Using by analogy the language of the correspondence
between the boundary ${ \cal N}=4$ four dimensional gauge theory and the bulk five dimensional
super-gravity (string theory) \cite{Mal}, by holography in this context we 
mean a correspondence between the loop equation, that we think as a theory
defined on boundary curves, and an equivalent effective action for the
eigenvalues of a Wilson loop, that we think as a theory defined on the bulk.  
The holographic form of the loop equation that we refer to is a preferred
form of the loop equation obtained by means of appropriate changes of 
variable so that this boundary-bulk correspondence becomes almost
manifest, for the reason that the quantum term in the loop equation, given as usual by
a contour integral along the loop, is computed as a regularized residue,
loop independent for self-avoiding loops.
This loop independence makes the holographic loop
equation close to admit an equivalent (holographic) effective action.
To this form of the loop equation we associate a "classical"
holographic effective action, $\Gamma$. $\Gamma$ is "classical" in the sense
that the loop equation for $\Gamma$ still contains a non-vanishing quantum
contribution given by the regularized residue. Yet, $\Gamma$ already
contains quantum corrections and indeed it is related to the quantum holographic
effective action, $\Gamma_q$, by means of a suitable conformal mapping and gauge fixing.
In fact the second ingredient needed to complete the holographic correspondence
is a conformal map of the region encircled by the based loop that
occurs in the loop equation to a cuspidal fundamental domain. On such domain
an effective action for the distribution of the eigenvalues of the Wilson loop,
$\Gamma_q$, can be indeed constructed, because the regularized residue vanishes.
Hence the holographic loop equation becomes equivalent to a critical
equation for $\Gamma_q$.
The second ingredient, i.e. the conformal map, is essentially new in this paper with
respect to \cite{MB} although it already appeared in implicit form in appendix C in 
\cite{MB}. Yet, there, the two-steps logic of holography and conformally
mapping was somehow mixed so that we did not realize that
the two dimensional technique of appendix C could be in fact extended to the
four dimensional case as well. 
More precisely, our quantum holographic effective action furnishes a critical equation
for the eigenvalues of the curvature of a unitary Abelian
local system \cite{S1,S2,S3,S4,Biq} in the
two dimensional case
and for the eigenvalues of the curvature of a non-unitary non-Abelian
twisted local system in the four dimensional case. By a twisted local system
we mean here a central extension of a (possibly infinite dimensional)
representation of the fundamental group
of a punctured Riemann surface. 
Thus, while the curvature of the local system determines directly the 
eigenvalues of the Wilson loop in $d=2$,
it does it only indirectly, via the non-Abelian gauge connection, in $d=4$. \\ 
To summarise, the basic ingredients of our holographic correspondence
are the following ones. \\
A holographic form of the loop equation in $d=2$ and $d=4$ as 
described in \cite{MB}. This form of the loop equation follows
from changes of variable, that though have a geometric meaning in terms of
symplectic reduction to a microcanonical ensemble, followed by a choice
of a holomorphic gauge, do not involve in a crucial way algebro-geometric
concepts. 
These are rather needed in the assignment of a local system with a lattice
of punctures in the $d=2$ theory
and of an infinite dimensional twisted local system with a lattice of 
punctures in the $d=4$ theory,
so that in both cases at least one of the punctures of such
systems lives at the
distinguished point of the based loop that enters the loop equation. \\
A conformal map of the region inside the based loop to a cuspidal
fundamental domain. \\
A regularization of the loop equation at the cusps, that amounts in fact to a
compactification at infinity of the cusps. \\
A reduction of the curvature
of the local system at the punctures to an Abelian sub-algebra in $d=2$ and to a Borel
sub-algebra in $d=4$, that follows by mapping conformally all the cusps (but one)
along the same line on the boundary of the upper-half plane, by
the choice of an axial gauge in a direction orthogonal to the boundary of the upper-half
plane and by the
compactification of the cusps at infinity. \\
It may be interesting to observe that the construction of the critical
equation for the quantum holographic effective action, which implies the holographic
loop equation, involves considering loops with a marked point taken away
and compactified at infinity after a change of the conformal structure.
From this point of view, the critical equation is the boundary theory
defined on the zero dimensional boundary of the loop (i.e. the marked point), while
the loop equation defines the theory on the bulk. 
The definition and the meaning of all these ingredients will be explained
in the following sections.
In section 2 we recall the holographic form of the loop equation
following \cite{MB}. This
section is needed to make this paper self-contained.
To check this form of the loop equation we reproduce
from it to lowest order in perturbation theory
the propagator in the $d=2$ case.
In section 3 we apply our algebro-geometric techniques to the
two dimensional case, in order to derive from the loop equation for 
self-avoiding loops the distribution of the eigenvalues that has been mentioned
as the motivation of this paper.
In section 4 we adapt the techniques of section 3 to the four dimensional case
and we compute the classical and quantum holographic effective action in terms of functional
determinants.
In section 5 we compute $\beta_0$, the first coefficient of the beta function,
from the four dimensional classical holographic
effective action, finding exact agreement with the perturbative result.
In section 6 we compute the contribution to $\beta_1$, the second coefficient of
the beta function, that arises from
the rescaling anomaly which occurs in passing from the Wilsonian to the
canonically normalised form of the holographic effective action.
This computation has some interest in itself and it is largely independent
of the entire holographic construction. It is also perhaps related to
a question about the beta function of $QCD$ raised in \cite{Shiff}.
In section 7 we collect some miscellaneous observations, referring especially
to analogies in the existing literature.
In section 8 we state our conclusions.

\section{The holographic loop equation}

We can summarise the basic philosophy in \cite{MB} as follows.
The loop equation in its usual form follows from the observation that the
integral of a derivative vanishes:
\bea
 0  =  \int DA_{\mu} Tr \frac{\delta}{\delta A_{\nu}(z)}
  (\exp(-\frac{N}{4 g^2} \int Tr F_{\mu \nu}^2 d^4x) \Psi(x,x;A))= \nonumber \\
  =  \int DA_{\mu} \exp(-\frac{N}{4 g^2} \int Tr F_{\mu \nu}^2 d^4x) 
  (Tr(\frac{N}{ g^2} D_{\mu} F_{\mu \nu}(z) \Psi(x,x;A))+ \nonumber \\
   +i\int_{C(x,x)} dy_{\nu} \delta^{(d)}(z-y) Tr(\lambda^a \Psi(x,y;A) \lambda^a 
  \Psi(y,x;A)) )
\eea  
where the sum over the indices $\mu,\nu$ in the action and over the
index $a$ is understood.
Here $ \lambda^a $ are Hermitian generators of the Lie algebra and 
\bea
\Psi(x,y;A)=P \exp i\int_{C_{(x,y)}} A_{\mu} dx_{\mu}
\eea
$\Psi(x,x;A)$ is the monodromy matrix of the connection $A_{\mu}$
along the closed loop $C(x,x)$ based at the point $x$, i.e. $\Psi(x,x;A)$ is
the Wilson loop.
For the group $SU(N)$ using the identity:
\bea
\lambda^a_{\alpha \beta} \lambda^a_{\gamma \delta}=
\delta_{\alpha \gamma} \delta_{\beta \delta}-
\frac{1}{N} \delta_{\alpha \beta} \delta_{\gamma \delta}
\eea
we get:
\bea
0=\int DA_{\mu}\exp(-\frac{N}{4 g^2} \int Tr F_{\mu \nu}^2 d^4x)
(Tr(\frac{N}{ g^2} D_{\mu} F_{\mu \nu}(z) \Psi(x,x;A))+ \nonumber \\
 +i\int_{C(x,x)} dy_{\nu} \delta^{(d)}(z-y) (Tr( \Psi(x,y;A)) Tr(\Psi(y,x;A))+ \nonumber
 \\ 
  -\frac{1}{N} Tr(\Psi(x,y;A)\Psi(y,x;A))))
\eea 
where the last term vanishes in the large-$N$ limit.
The first term is the classical contribution to the loop equation,
while the second term is the quantum contribution.
From  a modern point of view it is convenient to rephrase
the loop equation into an algebraic language \cite{Si}, since in this way a more
powerful view of what the problem is and of its difficulty is obtained.
Using the factorisation of gauge invariant operators \cite{Mig} in the large-$N$
limit and 
noticing that the expectation value can be combined with the matrix trace
to define a new generalised trace $ \tau $,
our problem is to find a (unique) solution $A_{\mu}(x)$ to
\bea
0=\frac{1}{ g^2}  \tau (D_{\mu} F_{\mu \nu}(z) \Psi(x,x;A)) + \nonumber \\
  +i\int_{C(x,x)} dy_{\nu} \delta^{(d)}(z-y) \tau( \Psi(x,y;A)) \tau(\Psi(y,x;A)) 
\eea
for every closed contour $C$, with values in a certain operator algebra with
normalised ($\tau(1)=1$) trace $\tau$ \cite{D,G,D1}. Such solution is named
the master field \cite{W}.
Thus the trace $\tau$ acts on a type $II_1$ von Neumann algebra \cite{Si}
generated by the monodromy operator $\Psi(x,x;A)$ of Wilson loops
based at $x$. This is the loop algebra, that is a representation 
of the homology algebra of based loops at $x$. Unfortunately,
even in the case in which $A_{\mu}(x)$ is Gaussian, this algebra is not
hyper-finite at $N=\infty$, that is, it is not the limit of a sequence of finite
dimensional matrix algebras, being a Cuntz algebra with an infinite
number of generators \cite{Haa1,Haa2,Cv}, which is algebraically isomorphic to
a free group factor with an infinite number of generators \cite{V}. 	
In the non-Gaussian case \cite{D,G,D1} there is no reason for which things
should be easier as to the hyper-finiteness property.
The point of view pursued in this paper is that we should abandon, 
at first stage, the idea of solving the loop equation defined over the entire loop algebra.
We should rather solve the following problem, whose solution
still conveys a lot of physical information, but that is 
algebraically considerably simpler.
For any fixed (self-avoiding, planar) based loop,
$C(x,x)$, we consider the von Neumann algebra generated by $\Psi(x,x;A)$ only,
which we want to determine from the loop equation restricted to $C(x,x)$.
This is a commutative von Neumann algebra, for which, as it is
well known, it there exists a structure theorem that is equivalent
to measure theory plus spectral theory of self-adjoint operators \cite{C}.
In particular every commutative von Neumann algebra is
type $I$ and thus hyper-finite \cite{Brown}. The trace on such algebra is a measure
determined by the distribution of the eigenvalues $ \rho_C(\lambda)$, counting multiplicity
\cite{C}:
\bea
\tau( \Psi(x,x;A))= \int \exp(i \lambda) \rho_C(\lambda) d \lambda
\eea
Thus our complicated algebraic problem reduces to finding the distribution
of eigenvalues for any given fixed (self-avoiding, planar) based loop $C(x,x)$. 
It should perhaps be repeated that now both the trace and the (commutative)
algebra are hyper-finite, so that the trace $\tau$, contrary to
the original problem, is completely known as the large-$N$ limit
of the normalised finite dimensional matrix trace $\frac{1}{N} Tr_N$.
In this respect we should mention that, already many years ago, 
the Migdal-Makeenko equation for any given loop was written in terms of the distribution
of the eigenvalues in a way that makes (implicit) use of the algebraic commutative structure 
associated to iterating a given loop \cite{Duh}.
For self-avoiding loops in $d=2$ in the weak coupling phase on a sphere
and for small $g^2 A$
the explicit answer for the distribution of the eigenvalues is the formula mentioned in the introduction.
In this paper we develop methods to solve for the distribution of the eigenvalues in
$d=2$ and $d=4$ directly from the loop equation for planar
self-avoiding loops.
As mentioned in the introduction our basic idea is holography: we would like
to map holographically the boundary loop equation into an equivalent bulk
holographic effective action, whose critical equation determines the distribution
of eigenvalues, $\rho_C$, via the eigenvalues of the curvature of a (twisted) local
system. It may be guessed that in doing so the Cauchy theorem will play a key role,
being a case (rather spectacular) of planar holography "ante litteram".
We manage to change variables in
the loop equation in such a way that in the new variables the quantum 
contribution be a regularized residue, loop independent for self-avoiding
loops. This has been achieved in \cite{MB}, changing variables in such a way
that functional differentiation of the monodromy in the new variables, that is the basic operation
to get the loop equation, produces the contour integral of a Cauchy kernel
instead of a delta-like contact term.
This makes us a first step closer to find an equivalent holographic quantum effective action. \\
The second step, by the way, following the $d=2$ analogy of a critical equation
that determines the distribution of eigenvalues, is to make the quantum contribution
vanishing in the large-$N$ limit. This is achieved in this paper
by a suitable conformal mapping: the key point is that evaluating the regularized
residue does not commute with the conformal mapping.
For our aim, we start with representing the partition function as an integral over microcanonical
strata, introducing a suitable resolution of identity \cite{MB}.
These strata are characterised by given levels of the curvature
of the gauge connection. 
Then we change variables to a holomorphic gauge in which the curvature $F$
acquires the form $F=\bar {\partial} (...)$. The new holographic loop equation
follows.
The last step, to annihilate the quantum term, is a conformal map
to a cuspidal fundamental domain.
We now describe our procedure in more detail,
proceeding in parallel in $d=2$ and $d=4$.
The $d=4$ case is not
substantially more complicated than the $d=2$ case from a purely
holographic point of view. 
As a first step we would like to change variable in the functional integral
from the gauge connection to the curvature. 
Formally this is done by means of the resolution of identity:
\bea
1=\int D\mu  
 \delta(F_A-\mu) 
\eea
in $d=2$ and:
\bea
1=\int D\mu_{\mu \nu}   
 \delta(F_{\mu \nu}(A)-\mu_{\mu \nu}) 
\eea
in $d=4$,
that we prefer to write decomposing the curvature into its anti-selfdual ($ASD$) and self-dual
($SD$) parts:
\bea
1=\int D\mu^-_{\mu \nu}   
 \delta(F^-_{\mu \nu}(A)-\mu^-_{\mu \nu}) \times \nonumber \\
\times \int D\mu^+_{\mu \nu}   
 \delta(F^+_{\mu \nu}(A)-\mu^+_{\mu \nu}) 
\eea
In \cite{MB} the resolution of identity associated to the $ASD$ constraint only was
imposed for reasons that will be cleared in section 4. 
The $SD$ and $ASD$ constraints are written in two dimensional language as Hitchin
equations \cite{Hi,Hi1}.
In particular the $ASD$ constraint is interpreted as an equation for the curvature of the
non-Hermitian connection $B=A+D=(A_z+D_u) dz+(A_{\bar z}+ D_{\bar u})
 d \bar z$ and a harmonic condition for the Higgs field 
$\Psi=-iD=-i( D_u dz+D_{\bar u}d \bar z )$.
$A$ is the projection of the four dimensional Hermitian connection
onto the  $(z=x_0+ix_1,\bar z=x_0-ix_1)$ plane of the planar loop
and $D$ is the projection of the four dimensional anti-Hermitian covariant derivative onto
the orthogonal $(u=x_2+ix_3,\bar u=
x_2-ix_3)$ plane.
In this paper we choose the following notation as far as the complex basis
of differentials $dz=dx_0+i dx_1$ and derivatives $\partial=\frac{\partial}{\partial
z}=\frac{1}{2}(\frac{\partial}{\partial x_0}-i\frac{\partial}{\partial x_1})$ is concerned.
Thus, for example, $A_z=\frac{1}{2}(A_0-iA_1)$.
We should notice that the observable in our loop equation is somehow adapted to the
microcanonical resolution of identity and thus in $d=4$ the Wilson loop involving the
non-Hermitian connection, $B$, is considered.
Yet, since at the end we would like to compute planar Wilson loops for
the Hermitian connection $A$, a more general kind of observables
is needed. Indeed in section 4 we will consider $B^{\lambda}=A+\lambda D$
in the limit $\lambda \rightarrow 0$, which we refer to as the unitary limit.
In this section for simplicity we limit ourselves to the first case.
We assume a partial Eguchi-Kawai reduction from four 
\cite{EK,Neu,Twc,Twl1,Twl2,RT}
to two dimensions \cite{MB}, that implies a rescaling of the classical action by a 
factor of $N_2^{-1}$ \cite{RT}. This factor is needed because the
partial Eguchi-Kawai reduction re-absorbs some space-time degrees of freedom
into the colour degrees of freedom. In our case a two dimensional torus is reduced to a point.
If the torus is commutative, $N_2$ is given by $\frac{1}{(2 \pi)^2}
 \int d^2x d^2p=
\frac{\Lambda^2 L^2}{( 2 \pi)^2}$. In the non-commutative case, instead, 
$N_2= Tr(1)=\sum_{n \leq N_2}1$.
The loop equation for $B$, in the partially reduced $d=4$ theory, reads:
\bea
0=\int DB_{\alpha} \exp(- \frac{N}{4 g^2}S_{YM})
(Tr(\frac{N}{4 g^2}\frac{\delta S_{YM}}{\delta B_{\alpha}(z)} \Psi(x,x;B))+ \nonumber \\
 - i\int_{C(x,x)} dy_{\alpha} \delta^{(2)}(z-y) Tr( \Psi(x,y;B)) Tr(\Psi(y,x;B)))
\eea 
with $\alpha=1,2$,
so that for these variables it is as difficult to solve as for the original
four dimensional connection $A_{\mu}$. 
After implementing in the functional integral the resolution of identity by
means of the gauge orbits of the microcanonical ensemble mentioned before,
we change variable in the loop equation from the connection
to the corresponding curvature in the holomorphic gauge. 
Thus we choose for the connection $A=a_z dz+\bar{a}_{\bar z} d{\bar z}$ in $d=2$ and
$B=b_z dz+ \bar{b}_{\bar z} d{\bar z}$ in $d=4$
the gauge $\bar {a}_{\bar z}$=0 and $ \bar {b}_{\bar z}$=0 respectively,
performing a gauge transformation in the complexification of the gauge group. 
This last change of variable is made in order to compute explicitly
the quantum contribution as a (regularized) residue.
It is well known by the Cauchy formula that the line integral along
a closed loop of a holomorphic
function times the Cauchy kernel with pole not lying on the loop depends
only on the winding number of the loop around the pole
and on the value of the holomorphic function at the pole of the Cauchy kernel:
\bea
f(z) Ind_{C}(z)= \frac{1}{2 \pi i} \int_C \frac{f(w)}{z-w}dw
\eea
where $Ind_{C}(z)$ is the winding number of $C$ around $z$.
The change of variables to the holomorphic gauge implies that
the Cauchy kernel is generated by functionally differentiating 
the monodromy of the connection in the loop equation with respect to the
integration variables in the functional integral. 
However, even if we reduce the computation of the quantum contribution
to the evaluation of a line integral of a Cauchy kernel, there are two
obstructions for the residue theorem to apply. The first one is that the
monodromy is not a holomorphic function of its endpoints. The second one is that
gauge invariance requires that the functional derivative with respect to the integration
variable be taken at a point
of the loop, for the result to be non-vanishing, i.e at a singularity
point of the Cauchy kernel.  
In fact we are going to compute a regularized residue obtained
as the line integral of a distribution.
This solves the regularity problem and the holomorphic problem
at the same time.
In $d=4$ for $B=A+i\Psi$ the 
microcanonical representation holds in the form:
\bea
Z=\int 
\delta(F_B-\mu) 
\delta(d^*_A \Psi- \nu) 
\exp(-\frac{N}{4 g^2}S_{YM}) DB D \mu D \nu 
\eea
where we have re-casted the $ASD$ microcanonical resolution of identity in
the Hitchin form of a curvature equation for $B$ plus a harmonic
condition for the Higgs field $\Psi$, for later convenience.
The curvature, $F_B$, is not Hermitian in general and thus we may include
in the functional integral the resolution of identity for its adjoint, in such a way that the 
integration measure is $D \mu D \bar{\mu}$ instead of $D \mu$.
Yet, since the monodromy of $B$ is in fact a functional of $\mu$ only, 
omitting $D \bar{\mu} $ does not affect the loop equation for $B$.
Our convention is that $\mu=\mu^0+n-\bar n$ is the curvature of $B$
in the basis $dx_0 \wedge dx_1$ with $\nu=n+\bar n$. Thus $\mu^0$ is Hermitian in this basis.
Only if we choose a basis different from the real one, indices are
added to $\mu$ to make the choice of basis explicit. From now on in this section
we mention only the $d=4$ case, since the $d=2$ case is obtained trivially
substituting the connection $B$ with the connection $A$ in $d=2$.
We now change variables to the holomorphic gauge,
in which the curvature of $B$ is given by the field $\mu'$,
obtained from the equation:
\bea 
F_B-\mu=0 
\eea
by means of a complexified gauge transformation $G(x;B)$ that puts the connection
$B=b+\bar b$ in the holomorphic gauge $\bar b=0$:
\bea 
\bar{\partial}b_z=-i\frac{\mu'}{2}
\eea
where $\mu'=G \mu G^{-1}$.
The partition function  becomes:
\bea
Z=\int 
\delta(F_B-\mu) 
\delta(d^*_A \Psi-\nu) 
 \exp(-\frac{N}{4g^2}S_{YM}) DB  \frac{D \mu}{D\mu'} 
D \mu'  D \nu
\eea
The integral over $B$ can now be performed and the resulting functional
determinants, together with the Jacobian of the change of variables to the
holomorphic gauge, absorbed into the definition of $\Gamma$.
$\Gamma$ plays here the role of a "classical" action, since we must
integrate still over the field $\mu'$. We may call $\Gamma$ the "classical"
holographic action, as opposed to the quantum holographic effective action,
to be found in section 4.
$\Gamma$ is written explicitly in section 4 in terms
of functional determinants.
The partition function is now:
\bea
Z=\int \exp(-\Gamma) D\mu' 
\eea
To realize our aim of getting the Cauchy kernel it is convenient
to study the loop equation for the Wilson loop involving the connection
$b$, thought as a functional of $B$ corresponding to gauge transforming $B$ into
the gauge $\bar b=0$. Such a gauge transformation belongs to the
complexification of the gauge group and it is rather a change of variable than
a proper gauge transformation. However, because of the property of the trace,
for closed loops, it preserves the trace of the monodromy. This allows us
to transform the loop equation thus
obtained into an equation for the monodromy of $B$. In our derivation of the loop 
equation, a crucial role is played by the condition that the expectation value
of an open loop vanishes.
In \cite{MB} two slightly different ways of achieving the vanishing of the
expectation value of open $b$ loops were presented. We may thus derive our
loop equation:
\bea
0=\int  D\mu'  Tr \frac{\delta}{\delta \mu'(w)}
  (\exp(- \Gamma)
  \Psi(x,x;b))= \nonumber \\
 = \int  D\mu' \exp(-\Gamma)
  (Tr(\frac{\delta \Gamma}{\delta \mu'(w)} \Psi(x,x;b))+ 
  \nonumber \\ 
  -\int_{C(x,x)} dy_z \frac{1}{2} \bar{\partial}^{-1}(w-y) 
  Tr(\lambda^a \Psi(x,y;b)
  \lambda^a \Psi(y,x;b)) )= \nonumber \\
 =\int D\mu' \exp(-\Gamma)
  (Tr(\frac{\delta \Gamma}{\delta \mu'(w)} \Psi(x,x;b))+ \nonumber \\
  - \int_{C(x,x)} dy_z \frac{1}{2} \bar{\partial}^{-1}(w-y)(Tr( \Psi(x,y;b))
    Tr(\Psi(y,x;b))+ \nonumber \\
  - \frac{1}{N} Tr( \Psi(x,y;b) \Psi(y,x;b))))
\eea
that in the large-$N$ limit reduces to:
\bea
0=\int D\mu'  \exp(-\Gamma)
  (Tr(\frac{\delta \Gamma}{\delta \mu'(w)} \Psi(x,x;b))+ \nonumber \\
  - \int_{C(x,x)} dy_z \frac{1}{2} \bar{\partial}^{-1}(w-y)Tr( \Psi(x,y;b))
    Tr(\Psi(y,x;b))) 
\eea
The only non-trivial case is when $w$ lies on the loop $C$. In this case
the loop equation can be transformed easily into an equation for $B$.
It is clear that the contour integration in the quantum term of the loop equation
includes the pole of the Cauchy
kernel. We need therefore a gauge invariant regularization.
We proposed in \cite{MB} several slightly different ways of regularising
the Cauchy kernel, that are essentially equivalent concerning the holographic form
of the loop equation.
The first one consists in analytically continuing the loop equation
from Euclidian to Minkowskian space-time. Thus $z \rightarrow i(x_+ + i \epsilon)$.
This regularization has the great virtue of being manifestly gauge invariant.
In addition this regularization is not loop dependent.
A natural, but loop dependent, regularization of the quantum contribution to the loop equation
is a $i \epsilon$ regularization of the Cauchy kernel in a direction normal
to the loop. We have the two possibilities of taking the internal or the
external normal. We check at the end of this
section that the solution of the loop equation does not depend indeed on this choice to lowest
order in perturbation theory in the $d=2$ theory.
Alternatively, as suggested in \cite{MB}, we may perform a conformal mapping of the region encircled
by the Wilson loop to the upper-half plane, followed by regularization
by means of analytic continuation to Minkowskian space-time or $i \epsilon$ regularization in a direction normal
to the loop.
A more sophisticated form of this regularization will be presented
in the next section, where we will map conformally the region whose boundary is the loop
with a marked point to a cuspidal fundamental domain.
Quite interestingly we will find that the regularization of the loop equation does
not commute with this conformal mapping. Indeed we will take advantage of this fact
to annihilate the quantum term, after having accounted for the effects of the
conformal map everywhere else. 
In any case the result of the $i \epsilon$ regularization of the Cauchy kernel is the sum of
two distributions, the principal part plus
a one dimensional delta function: 
\bea
\frac{1}{2}\bar{\partial}^{-1}(w_x -y_x +i\epsilon)= (2 \pi)^{-1} (P(w_x -y_x)^{-1}
- i \pi \delta(w_x -y_x))
\eea
The loop equation thus regularized looks like:
\bea
 0=\int D\mu' \exp(-\Gamma) 
  (Tr(\frac{\delta \Gamma}{\delta \mu'(w)} \Psi(x,x;b))+ \nonumber \\
   - \int_{C(x,x)} dy_x(2 \pi )^{-1} (P(w_x -y_x)^{-1}
  - i \pi \delta(w_x -y_x)) \times \nonumber \\
  \times Tr( \Psi(x,y;b)) Tr(\Psi(y,x;b)))
\eea
Being supported on open loops the principal part does not contribute and the
loop equation reduces to:
\bea
0=\int D\mu' \exp(-\Gamma)
  (Tr(\frac{\delta \Gamma}{\delta \mu'(w)} \Psi(x,x;b))+ \nonumber \\
   +\int_{C(x,x)} dy_{x}
\frac {i}{2}\delta(w_x -y_x) Tr( \Psi(x,y;b)) Tr(\Psi(y,x;b)))
\eea
Taking $w=x$ and using the transformation properties of the $b$ monodromy
and of $\mu(x)'$, the preceding equation can be rewritten in terms
of the connection, $B$, and the curvature, $\mu$:
\bea
0=\int D\mu' \exp(-\Gamma)
  (Tr(\frac{\delta \Gamma}{\delta \mu(x)} \Psi(x,x;B))+\nonumber \\
   +\int_{C(x,x)} dy_{x}
\frac {i}{2}\delta(x_x -y_x) Tr( \Psi(x,y;B)) Tr(\Psi(y,x;B)))
\eea
where we have used the condition that the trace of open loops vanishes
to substitute the $b$ monodromy with the $B$ monodromy.
This is our final form of the regularized Euclidian loop equation (there is an
analogous form in Minkowskian space-time).
Let us notice the sign ambiguity in the quantum contribution
that depends on the choice of the sign of $ i \epsilon$.
Because of the product of traces, in general the quantum contribution does
not have the
same operator structure as the classical term. In addition the quantum contribution
is loop dependent in general.
However, for self-avoiding loops, the quantum contribution
is just a linear topological term added to the "classical" action, $\Gamma$,
provided the loop equation is interpreted in a strong sense, that is as:
\bea
0=\frac{\delta}{\delta \mu(x)}(\Gamma + \frac{i}{2}\int d^2x Tr\mu)
\eea
An interpretation of this kind was attempted in \cite{MB}, where the central
term, being topological, was thought to be essentially irrelevant or cancelled
against an anomalous phase in the effective action.
Unfortunately, following the interpretation of \cite{MB}, we were
unable to reproduce to lowest order in perturbation theory
the gauge propagator.
The reason is that the central term is absolutely essential to obtain 
the gauge propagator to lowest order from the loop equation for self-avoiding
loops, if the loop equation is interpreted weakly as:
\bea
0=\tau((\frac{\delta \Gamma}{\delta \mu(x)}+\frac{i}{2}Tr(1))\Psi(x,x;B))
\eea
In fact the role of the central term in getting the correct propagator
is entirely analogous to the role of the contact term in the original
loop equation (Eq.(5)). Hence we might conclude that we have come to a loose end
despite our sophisticated changes of variable: perhaps we have 
simply rewritten our loop equation in an exotic way for exotic
variables. In the two next sections we will see that, thanks to some
new ingredient, this is not in fact the case. \\
We end this section checking to lowest order that the gauge propagator
does follow from the loop equation in the holographic form in $d=2$,
when it is interpreted in a weak sense.
For computational convenience we consider the holographic loop equation in $d=2$
followed by Eguchi-Kawai reduction. We interpret it in a weak sense
and we expand in powers of $\mu(x)=\exp(ipx) \mu(0) \exp(-ipx)$.
We get for a small loop:
\bea
\tau(\frac{L^2}{2g^2 N_2} \mu(0)^2 -\frac{1}{2})=0
\eea
where the factor of $N_2^{-1}=N^{-1}$ takes into account the Eguchi-Kawai reduction
from $d=2$ to $d=0$ and $L^2$ is the area of the space-time torus.
A somehow intriguing feature of our computation, based on the
Eguchi-Kawai reduction, is that we perform it
in a way that implies (to lowest order) 
a hyper-finite trace. Yet this does not imply necessarily hyper-finiteness
of the algebra, that would follow instead from uniform hyper-finiteness of the
trace \cite{Brown}. Indeed even for the solution in terms of free random variables in $d=2$,
that certainly generate a non-hyperfinite algebra, a solution
admitting a hyper-finite trace has been proposed \cite{G}.
Thus if we set:
\bea
\tau= \lim_N \frac{1}{N} Tr_N
\eea
we get for $\mu$ from Eq.(26) the normalisation:
\bea
\mu_{ik}(x)= \frac{g}{L} \exp(i(p_i-p_k)x)
\eea
for $i > k$ with $\mu_{ii}=0$ and $\mu=\mu^*$.
Since from Eq.(8) it follows, to lowest order in powers of $\mu$,
in the gauge $D_{\alpha}\delta A_{\alpha}=0$:
\bea
A_z= i \partial (-\Delta)^{-1} \mu \nonumber \\
A_{\bar z}=- i \bar \partial (-\Delta)^{-1} \mu
\eea
we get for the propagator to lowest order:
\bea
\tau (A_z(x)A_{\bar z}(y))=\frac{g^2}{4 N L^2} \sum_{i \neq k}(\frac{1}{(p_i-p_k)^2}
\exp(i(p_i-p_k)(x-y))) \sim \nonumber \\
\sim \frac{g^2}{4(2 \pi)^2} \int \frac{1}{p^2} \exp(ip(x-y)) d^2p
\eea
as it should be in the quenched theory.

 \section{The two dimensional case: reduction to an Abelian sub-algebra}

The aim of this section is to solve the holographic loop equation of the
previous section by means of a certain Abelianization map in the two dimensional
case. 
The advantage of considering the two dimensional case first is that
in two dimensions we already know the exact answer, so that we can test
our ideas about solving the loop equation for self-avoiding loops.
As mentioned in the introduction, to realize the Abelianization map
we need:\\
a local system; \\
a uniformization to a cuspidal fundamental domain; \\
a regularization at the boundary cusps that amounts to a compactification of the
cusps at infinity; \\
a clever choice of the gauge. \\
The local system was introduced in \cite{MB} basically to have a
mathematically well defined model of the moduli space of the master field.
Yet, it turns out that the need of a local system and in particular of the
associated singularities has a deeper meaning.
The local system furnishes a lattice or adelic interpretation of the 
microcanonical localisation:
\bea
F_A= \sum_p \mu_p \delta^2(x-x_p)
\eea
in the  holographic loop equation:
\bea
0=\int D\mu'  \exp(-\Gamma)
  (Tr(\frac{\delta \Gamma}{\delta \mu(w)} \Psi(x,x;A)) +\nonumber \\
  - \int_{C(x,x)} dy_z \frac{1}{2} \bar{\partial}^{-1}(w-y) Tr( \Psi(x,y;A))
  Tr( \Psi(x,y;A)))
\eea
From the point of view of the functional integration the curvature
of the connections associated to local systems is dense in the sense 
of distributions in the space of curvatures.
For local systems the loop equation acquires the following form:
\bea
0=\int \prod_q D \mu_q '  \exp(-\Gamma)
  (Tr( \frac{ \delta \Gamma}{ \delta \mu_p} \Psi(x_p,x_p;A)) + \nonumber \\
 -  \int_{C(x_p,x_p)} dy_z \frac{1}{2} { \bar{ \partial}}^{-1}(x_p-y)Tr( \Psi(x_p,y;A))
    Tr( \Psi(y,x_p;A)))
\eea
where now it is necessary that the marked point of the based loop
coincides with a puncture to get
a non-trivial loop equation.
It can be checked by direct computation that integrating in the 
functional integral over non-Abelian local systems leads the same result
for the Wilson loop as in the lattice theory:
\bea
\tau( \Psi(A;x,x))= Z^{-1} \int \prod_p dg_p
Tr(g_p \exp(i \theta_p) g_p^{-1}) \times \nonumber \\
\times \exp(- \sum_p \frac{N}{ 2 g^2 a^2 }
Tr( \theta_p^2)) \prod_p \prod_{i \ne j} |\exp(i \theta^i_p) - \exp(i \theta^j_p)| d \theta_p  
\eea
where the product over $p$ is restricted to the lattice points internal to the loop
and $a^{-2}=\frac{\Lambda}{(2 \pi)^2}= \delta^{(2)}(0)$.
The $ dg_p $ integrals can be easily performed leading to the correct
lattice result \cite{GW} (the only difference is the
Wilsonian lattice action instead of its formal continuum limit) :
\bea
\tau( \Psi(A;x,x))= Z^{-1}\prod_p \int (Tr( \exp(i \theta_p )) \exp(- \frac{N}{2 g^2 a^2} Tr(\theta_p^2))
 \times \nonumber \\
 \times \prod_{i \ne j} |\exp(i \theta^i_p) - \exp(i \theta^j_p)| d \theta_p \nonumber
\eea
that is equivalent for $ \tau( \Psi) $ to the formula for the distribution of
eigenvalues in the introduction for small $g^2 A$ and
in the scaling limit $N_C \rightarrow \infty $ with $N_C a^2=A=constant$,
where $N_C$ is the number of punctures inside the loop $C$.
We will see now how this result is reproduced by the Abelianization map
of the holographic loop equation.
The key point is that identifying the marked point of the based loop
with a puncture contains implicitly the possibility
of a change of the conformal structure. 
Formally the Wilson loop is invariant under re-parameterisation of the boundary that can be 
extended to conformal transformations of the region encircled by the loop,
but in fact we will change the conformal structure in a way that is equivalent
to attach to the loop, in a neighbourhood of $x$, an infinitesimal strip going to infinity.
This does not alter the Wilson loop because of the zig-zag symmetry \cite{Pol2}.
Of course the classical action
in $d=2$ is not conformally invariant, so that if we are going to compute
the classical action in terms of a conformally transformed local system
we must transform the classical action properly.
Coming back to the loop equation for local systems, it is of the utmost importance that the regularized 
residue with the new conformal structure vanishes in the loop
equation. We now explain why, first proceeding heuristically and then by
direct computation.
Our lattice gives rise to a punctured sphere with a based 
contour, $C(x,x)$, that determines an internal region
(the smaller one on a large sphere), $\Omega_x$, with the topology
of a disk
with some punctures inside and at least one on the boundary, that coincides with the
distinguished base point of the loop, $x$. $\Omega_x$ can be mapped conformally to a cuspidal
fundamental domain on the upper-half plane with all the cusps but one, for example
the cusp on the boundary,
on the $y=0$ axis and the remaining one at $y= \infty$. On the other hand also the region
external to the loop can be mapped conformally in a similar way.
Heuristically this conformal map
allows us to diagonalise the curvature at the cusps, thanks
to a residual gauge symmetry (see below), provided the cusps
are compactified. In general the compactification is not possible,
otherwise every local system on a sphere would be Abelian.
However, in our case, the compactification is required since the marked point, $x$,
in origin belongs to the closed loop $C(x,x)$ and thus it is not a puncture.
In addition, assuming rotational invariance on the sphere, if one cusp
is compactified also all the remaining ones should be.
On the fundamental domain all the cusps but one 
are on the $x$-axis and hence, choosing an axial gauge along the $y$-axis, there is a residual
gauge symmetry of making gauge transformations along the $x$-axis to
diagonalise the curvature at the cusps on the $x$-axis. In addition, since our
local systems define a representation of the fundamental group
of a punctured sphere, also the curvature at the remaining cusp at $y=\infty$ must be diagonal.
Thus we get an Abelian local system.
But Abelianization can only be consistent with a vanishing of the
regularized residue in the loop equation, since an Abelian system becomes classical in the large-$N$
limit. Thus there is no quantum term in the large-$N$ loop equation on the
cuspidal fundamental domain, and
being gauge invariant, the loop equation is to coincide
before and after gauge fixing. Hence there never is a quantum residue in any gauge.
Notice that the Abelianization could not be performed before
the conformal mapping, since it is essential that the cusps
be on the same line and compactified.
Now we must check directly that the quantum residue vanishes in any gauge.
The cuspidal domain is a polygon defined by the uniformization
theory of Riemann surfaces with punctures and boundaries.
A unified approach in which the boundaries are treated in a way similar 
to the punctures has been given by Penner \cite{Pen}. Remarkably
Penner approach involves the choice of at least one puncture on the
boundaries, precisely as required by our interpretation of the holographic loop
equation in terms of local systems.
The boundary arcs of the polygon that uniformizes our region $\Omega_x$
are oriented in the following way.
The couple of arcs that end into a cusp corresponding to the punctures in the
interior of $\Omega_x$ have opposite orientations on the fundamental domain,
since they are in fact
identified by the gluing map that reconstructs from the polygon
the punctured Riemann surface: this identification creates tubes out of
such cusps. As a consequence these couple of arcs share the same orientation
on the reconstructed Riemann surface.
On the contrary, the arcs ending into the cusp lying on the loop share the same
orientation on the polygon since they are not glued together:
these arcs are associated to an infinitesimal strip going to infinity.
As a consequence these couple of arcs have opposite orientation on the
strip to infinity.
This difference in orientation plays a crucial role in
evaluating the regularized residue at the internal cusps and at the boundary cusps.
In fact in the first case the regularized residues associated to the two asymptotes of the cusp
sum up to $1$ because of the same orientation of the asymptotes on the 
Riemann surface.
In the second case
the sum is $0$, because the opposite orientation of the asymptotes on the
Riemann surface. We may look at the last fact as just
another consequence of the zig-zag symmetry.
Thus the quantum contribution vanishes. 
We can summarise our argument as follows.
The local system is invariant under the conformal map,
while the classical action transforms in a definite way.
The marked point of the based loop becomes a cusp of the
fundamental domain. This cusp is obtained adding an infinitesimal strip going to infinity
to the loop in a neighbourhood of the marked point.
Because of the zig-zag symmetry, at this cusp the quantum regularized
residue vanishes.
We can take into account the preceding argument to derive a new form
of the holographic loop equation
that leads to a critical equation for a holographic quantum effective action.
The resolution of the identity involves now local systems on the
cuspidal fundamental domain with coordinates $(t, \bar t)$:
\bea
1= \int \delta(F^{(t)}_A-\mu^{(t)}) D\mu^{(t)}
\eea
where the superscript $^{(t)}$ refers to fields defined on the fundamental domain.
It is this new resolution that is inserted into the functional integral:
\bea
 0  =  \int DA D\mu^{(t)}
  \exp(-\frac{N}{2 g^2} \int Tr F_{A}^2 d^2x) \delta(F^{(t)}_A-\mu^{(t)}) D\mu^{(t)}
\eea  
Then all the steps go through as in the second section. The universal
cover of the fundamental domain is the upper-half plane $U$.
On $U$ we choose the gauge
$A_y=0$, that leaves, as a residual gauge symmetry, gauge transformations that are
$y$ independent. In the gauge $A_y=0$ the determinant due to localisation,
i.e. the one obtained integrating with respect to the gauge connection,
$A$, the delta functional in Eq.(36),
and the Faddeev-Popov determinant are both field independent and
cancel each other. We can use the residual gauge symmetry to fix the gauge
$\mu_{p}^{ch}=0$ (the label $ch$ means the non-diagonal part) at the cusps
on the $x$-axis. The associated extra Faddeev-Popov determinant is the square of the Vandermonde
determinant of the eigenvalues of the curvature of the local system at the punctures. 
Then the holographic quantum effective action reduces to:
\bea
\Gamma_{q}= \frac{N}{2 g^2 a^2} \sum_i \sum_p |\frac{\partial t}{\partial z}(p)|^2
 h^{(t)i 2}_p - \sum_{i \ne j} \sum_p log|h^{(t)i }_p -h^{(t)j }_p | + \nonumber \\ 
 - logDet(\frac{Dh^{(t)}}{Dh^{(t)'}})
\eea
where the $^{(t)}$ superscript refers to the domain of definition
of the lattice field, $h_p$, and we have set $a=\frac{2 \pi}{\Lambda}$, 
with $a$ the lattice spacing
corresponding to the cutoff $\Lambda$ of the theory, that comes from the product of delta
functions at the same point in the classical action. The last term is the logarithm of the 
Jacobian to the holomorphic gauge of the Abelian local system. However it vanishes identically
in a gauge in which the curvature is Abelian (see at the end of next section).
It should be noticed that $\Gamma_{q}$ is expressed
as a functional of the local system on the fundamental domain. This
involves in the classical term a change of the metric since the classical action is not
conformally invariant.
Since on the fundamental domain the quantum term vanishes, the loop
equation for self-avoiding loops reduces to:
\bea
0=\tau(\frac{\delta \Gamma_{q}}{\delta h^{(t)}_p}\Psi(x_p,x_p;A)) 
\eea
Eq.(38) is implied by the critical equation:
\bea
\frac{\delta \Gamma_{q}}{\delta h^{(t)}_p}=0
\eea
Getting a saddle-point
morally implies that we have somehow integrated away the order
of $N^2$ non-Abelian degrees of freedom to obtain the effective action
for the order of $N$ remaining eigenvalues. It is the conformal map
that allowed us to diagonalise the curvature at the cusps.
But this works only for a system in which the curvature has delta-like singularities.
Hence local systems know about quantum field theory ! It is perhaps this
the very reason for the occurrence of Hitchin systems in $d=4$ 
quantum field theories (see section 4 and section 7).
Notice that the Abelianization works only for self-avoiding loops.
In case of self-intersection the marked point would be ramified
and globally there would not be enough residual symmetry
in an axial gauge on a ramified covering to diagonalise all the cusps.
On the other hand the regularized residue in the loop equation does not vanish in
general at a ramification point.
It remains to be seen that our critical equation coincides with the large-$N$
saddle point in Eq.(1).
To do so we need a better understanding of the uniformization map.
It turns out that the relevant mathematics is still the one of the moduli
space of bordered Riemann surfaces following \cite{Pen}.
As already noticed,
Penner version of the Teichmuller theory of bordered surfaces
involves the choice of at least a puncture and a horocycle arc in a neighbourhood
of the puncture on each boundary of the surface, in analogy with the theory
of punctured surfaces that involves a horocycle around each internal puncture.
More generally we may consider the case in which the loop
in the loop equation intersects a number of punctures of our lattice.
It is clear that the preceding arguments about the vanishing of the quantum
residue apply in this more general situation, since they depend indeed on
the local structure around each puncture. But now, since all the punctures on the loop
are attached to infinitesimal strips going to infinity, the circle
at infinity contains a lattice (or adelic) image of the original loop
through the conformal map. 
Quadratic differentials \cite{Pen1} can be used to construct the uniformization
map from the generic region that occurs in the loop equation
to the cuspidal fundamental domain.
The basic relation between quadratic differentials, $q$, and the uniformization map
is:
\bea
\frac{\partial t}{\partial z}= \sqrt q
\eea
We need therefore the standard form of a quadratic differential
near a cusp:
\bea
\frac{\partial t}{\partial z}=  \frac{L}{2 \pi i z}
\eea
where $L$ is the length of the horocycle arc around the cusp.
Since this expression is infinite at the cusps it must be regularized and suitably
interpreted.
In particular it depends crucially on what the cutoff is on the fundamental
domain near the cusps. This may be difficult to understand in general,
but it seems easier in the case of a circular loop. In this case we have essentially a circle
that is mapped (adelically) into the circle at infinity. This is a cylinder, i.e. a punctured disk,
which is mapped by the uniformization map to a strip in the upper-half plane.
In this case the uniformization map is:
\bea
t= \frac{L}{2 \pi i }log(z)
\eea
Thus we get:
\bea
|\frac{\partial t}{\partial z}|(p)^2= \frac{R^2}{a^2}=\frac{A}{ \pi a^2} \sim N_C
\eea
where $R$ is the radius and $A$ the area of the disk, while $a$ is the radius
of a little disk around the puncture. Thus $N_C$ is the number of lattice
points inside the disk.
When Eq.(43) is inserted into Eq.(37) the correct distribution of eigenvalues,
given by the saddle point equation for the effective action in Eq.(1), is obtained after noticing that
$ \theta^i $ in Eq.(1) is related
to $h^i_p=h^i$ (by rotational invariance on the sphere) in Eq.(37) by:
\bea
\theta^i=N_C h^i
\eea
since the magnetic flux through the loop counts the number, $N_C$,
of punctures inside. In fact the holographic action in terms of $ \theta^i $
becomes $\sum_p (\frac{N}{2 g^2 N_Ca^2} \sum_i \theta^{i 2}
-\sum_{i \ne j} log|\theta^{i } -\theta^{j}|+constant) $
which differs by an irrelevant overall factor and an additive constant from the one in
Eq.(1). 
In this paper we have ignored the contribution of the region external to the loop
and the corresponding conformal map. We will consider it elsewhere.
It should be noticed that the eigenvalues of the curvature at the
punctures of the local system are in fact defined modulo $ 2 \pi $.
We can take into account the periodicity of $h_p$ summing over appropriate
windings.
This is expected to lead to the known phase transition for $g^2 A$ large
in the large $N$ limit \cite{GW,D2,KA}, but in that case the external region
must be considered too.

\section{The four dimensional case: reduction to a Borel sub-algebra}

The key points in the previous section can be extended to the
planar four dimensional case. 
We list here the needed modifications. 
We should remind that in our approach the observable in the loop equation is adapted to the
microcanonical resolution of identity. In the four dimensional case the
resolution of identity involves the $SD$ or $ASD$ curvature.
Thus, for example in the $ASD$ sector, the connection that enters the loop equation is $B=A+D$,
which is non-unitary. However for physical reasons we are interested to compute
the Wilson loop for the unitary connection $A$.
Thus we introduce $B^{\lambda}=A+(\lambda D_u dz+ \bar \lambda D_{\bar u} d \bar z)$ 
where  $\lambda$ is a section
(possibly constant)
of a holomorphic line bundle and we take the limit $\lambda
\rightarrow 0$. We refer to this limit as the limit of unitary Wilson loop
or the unitary limit in short.
In the four dimensional case it is most convenient to take the space-time
to be a product of a two dimensional sphere by a non-commutative torus
in the limit of infinite non-commutativity \cite{MB}.
This limit is known to be equivalent to the usual commutative theory
in the large-$N$ limit \cite{Twc}. Although several different space-times may be considered,
in this paper we make this choice because the associated $SD$ or $ASD$
equations look like a kind of infinite dimensional vortex equations (see below).
Heuristically we find appealing the occurrence of vortices for the hope of
reproducing both the area law at large distances and the Coulomb law at short distances.
The choice of the sphere is also modelled on the analogy with the
two dimensional case, especially for simplifications due to the non-existence
of non-trivial cycles, but for the ones associated to the punctures.
Indeed on a sphere all the moduli of the local system are the local moduli.
When we take the unitary limit in our loop equation, some care
is necessary, since the limit has to be taken in such a way that the correct four
dimensional information survives, for example in the beta function.
The first coefficient of the beta function (see
next section) can be computed in two different but related ways, that we now explain.
Since the theory lives on a product of a sphere by a non-commutative torus,
the curvature equation involves a central term, $H$,
equal to the inverse of the parameter of non-commutativity,
$\theta$. This occurs because, once the gauge connection is required to vanish at infinity
up to gauge equivalence, the only term that survives in the curvature at infinity
is the commutator
of the derivatives on the non-commutative torus, that is $H$. In turn
$H$ vanishes as $\frac{1}{N}$ in the large-$N$ limit.
Therefore, in the case of $B^{\lambda}$, our centrally extended and $\lambda$ twisted
$ASD$ curvature equation reads:
\bea
F_A-i |\lambda |^2\Psi^2= \sum_{p} \mu^0_p \delta^{(2)}(x-x_p)+ H1\nonumber \\
\lambda \bar{\partial}_A \psi= \sum_{p} \lambda n_p \delta^{(2)}(x-x_p)\nonumber \\
\bar \lambda \partial_{A} \bar{\psi}=\sum_{p} \bar \lambda \bar{n}_p \delta^{(2)}(x-x_p)
\eea
where we have set $D=i \Psi$, we have rescaled $n$ by a factor of $\lambda$
to ensure the finiteness of $\psi$ in the $\lambda \rightarrow 0$ limit
and analogously in the $SD$ case.
The central extension $H$ is referred to in this paper as the twist of the local system.
The $\lambda$ rescaling instead is a twist of the Hitchin system 
considered as a twisted Higgs bundle.
In our interpretation of the preceding equations as vortex equations, the central extension
$H$ is related to the non-vanishing of the Higgs field at infinity, while the zeroes
of the Higgs field may arise from twisting by the factor of $\lambda$. In fact
this is precisely what we require in the unitary limit. We choose a $\lambda$
that has a lot of zeroes in a compact set containing the loop $C$, in order
to make the Higgs field vanishing small in a neighbourhood of the loop to ensure
unitarity of the monodromy along $C$,
and that converges
to $1$ at infinity, to keep the information about the four dimensional nature
of the theory.
As to the beta function, the simplest case is $\lambda=1$. In this case,
if we are interested only in the ultraviolet logarithmic
divergences for
the collective field $\mu^0$ and not in the critical equation for the holographic quantum
effective action,
we need not to decompose the collective field $\mu^0$ into a sum of delta distributions,
that is equivalent to introduce the local system, and in fact
$\beta_0$ can be found directly from the ultraviolet divergences of the "classical" holographic action,
$\Gamma$, looking at the $Tr(\mu^{02})$ counter-term (see next section), solving in perturbation theory,
at first order in power of the local curvature, the preceding equation around the $\mu^0 =n=0$
solution.
However our observable in the case $\lambda=1$ is not physically interesting. In addition
it is trivial at least at first order in perturbation theory and in the large-$N$ limit.
This occurs because the contribution of the propagator of the field $A_z$ is exactly cancelled
by the one of $A_u$ because of the different factors of $i$ in the Wilson loop.
In fact at $\lambda=1$ the Wilson loop is probably topological in the large-$N$ limit.
The limit $\lambda \rightarrow 0$ is the most physically interesting.
In this limit $H$ must be left untouched at infinity
since it is essential to keep the correct four dimensional information while $n$
is expected to be rescaled by a factor of $\lambda$. A little thought shows
that this may happen if $\lambda$ has a zero at each puncture in a compact set containing
the loop and converges to $1$ at
infinity as already anticipated.
This implies that $B^{\lambda}$ converges to
$(A_z+\partial_u)dz+(A_{\bar z}+\partial_{\bar u}) d \bar z $ at infinity and to 
$A$ on a compact set on the sphere in the complement of infinity.
In this case it is more convenient the lattice or adelic interpretation
of the curvature equation in order to obtain $\beta_0$.
It is natural to interpret the adelic theory as defining a scaling limit as in usual lattice
gauge theories. In such theories there are no ultraviolet
divergences because of the finite lattice spacing, but all the 
divergences appear as logarithmic infrared divergences at a scale much
larger than the lattice spacing but still smaller than the inverse
of $ \Lambda_{QCD} $, that goes to infinity in the scaling limit,
i.e. when the coupling constant goes to zero. If we are interested in the
infrared logarithmic divergence we should employ in our computations
the asymptotic value of $\lambda$ at infinity, i.e. $1$, and hence our
computation reduces to the one in the case $\lambda=1$ everywhere.
For our twisted local system the region of asymptotic freedom is thus
the large distance region on the sphere.
This perhaps resembles the holographic
UV-IR duality already encountered in \cite{Mal}.
From a technical point of view, as far as $\beta_0$ is concerned, the computation
for $\lambda=1$ and the one for $\lambda \rightarrow 0$ are identical once it is observed that
the in the second case we are actually looking at a neighbourhood of infinity. 
Thus in this paper we report only the first case in the next section.
We have seen that having introduced non-commutativity leads inevitably
to a central extension in the curvature equation and thus for $\lambda=1$
we get a central extension of an infinite dimensional representation of the
fundamental group
of the punctured Riemann surface. This central extension cannot
be decomposed in general into a $U(1)$ Abelian system and a flat connection as in finite
dimensions. Thus we may wonder which are the finite dimensional
approximations of this system. The answer is quiver bundles with
parabolic singularities \cite{Qui}.
In $d=4$ the additional difficulty arises that is not possible to impose
the two $SD$ and $ASD$ 
constraints independently at the same time for an irreducible connection in the functional
integral. 
Thus either we impose only the $ASD$ part of the resolution of identity
as we did in \cite{MB} (we are not finding a universal master
field, since our observable is adapted to the resolution of identity) or we introduce orbifold models as follows. 
There are orbifold models whose large-$N$ limit is equivalent to the
large-$N$ limit of $QCD$.
We explain what a orbifold is in this context following \cite{Orb}:" A certain
parent gauge theory
is chosen", in our case pure $SU(N)$ gauge theory; "the orbifold theory is simply
given throwing away
all fields that are not invariant under a discrete subgroup of the gauge symmetry
" in our case $Z_2$. "The resulting theory has the remarkable property that at
large-$N$ its perturbation series
is the same as the parent theory, up to some simple rescaling of $N$".
Sometimes, using the loop equation, it is even possible to show non-perturbative
equivalence
between the parent and orbifold theory.
This is the case for our $Z_2$ orbifold, that is simply a gauge theory with
gauge
group $SU(N) \times SU(N)$. It is clear that if we compute the Wilson loop in
the $(N, \bar N)$
representation in the $Z_2$ orbifold theory in the large-$N$ limit, this will be
the same as the Wilson loop in the adjoint representation
of the parent $SU(N)$ theory. 
As we said the necessity of a $Z_2$ orbifold occurs
in $d=4$, if we want to impose
in the functional integral the two $SD$ and $ASD$ constraints independently at the
same time.
We now write the loop equation in terms of the lattice field of curvatures 
in $d=4$:
\bea
0=\int \prod_q D\mu_q '  exp(-\Gamma)
  (Tr(\frac{\delta \Gamma}{ \delta \mu_p}\Psi(x_p,x_p;B)) + \nonumber \\
 -  \int_{C(x_p,x_p)} dy_z \frac{1}{2}\bar{\partial}^{-1}(x_p-y)Tr(\Psi(x_p,y;B))
    Tr(\Psi(y,x_p;B)))
\eea
and analogously in the unitary limit:
\bea
 Z=\int 
 \delta(F_{ B^{\lambda} }-\mu^{\lambda} - H 1) 
 \delta(d^{*}_{A^{\lambda}} {\Psi}^{\lambda}-\nu^{\lambda}) \times \nonumber \\
  \times \exp(-\frac{N}{4g^2}S_{YM}) DB  \frac {D \mu^{\lambda} }{D \mu'^{\lambda} } 
 D \mu'^{\lambda}   D \nu^{\lambda}=  \nonumber \\
=\int exp(-\Gamma^{\lambda}) D \mu'^{\lambda}  
\eea
where a $\lambda$ dependent effective action appropriate for the
study of the monodromy of the operator 
$B^{ \lambda }=A+i (\lambda \psi+\bar{\lambda} \bar{\psi})= A^{ \lambda }+
i {\Psi}^{\lambda} $, with $\mu_{z \bar z}^{\lambda}=\mu_{z \bar z}^0+
\lambda n_{z \bar z}- \bar \lambda \bar n_{z \bar z}$, must be introduced.
As in $d=2$, to get rid of the quantum term, we must map the lattice
loop equation to a cuspidal fundamental domain.
In doing so we make a conformal rescaling of the Hitchin system as dictated by
the conformal properties induced requiring the conformal invariance of the
non-unitary monodromy that enters the loop equation:
\bea
B_z= \frac{\partial t}{\partial z} B_t
\eea
Thus, in mapping to the fundamental domain, we are actually
rescaling in a certain way also $H$, the central extension, and the orthogonal coordinates
as well.
Now, to get a zero quantum contribution, we must insert in the functional integral
the Wilson loop and the resolution of identity,
expressed in terms of the Hitchin system on the fundamental domain (we do not
display explicitly the dependence on $\lambda$):
\bea
 Z=\int 
 \delta(F^{(t)}_{ B }-\mu^{(t)} - H^{(t)}1) 
 \delta((d^{*}_{A} {\Psi})^{(t)}-\nu^{(t)}) \times \nonumber \\
  \times \exp(-\frac{N}{4g^2}S_{YM}) DB  \frac {D \mu^{(t)}}{D \mu'^{(t)}} 
 D \mu'^{(t)}  D \nu^{(t)}=  \nonumber \\
=\int \exp(-\Gamma) D \mu'^{(t)}
\eea
As in $d=2$ the Wilson loop is invariant by construction, up to the added infinitesimal
strip to infinity, that is attached and got rid by the zig-zag symmetry.
The classical action must be expressed in terms of the Hitchin system on the fundamental
domain as in the $d=2$ theory:
\bea
S_{YM}= \frac{1}{N_2} \int \frac {d^2z}{4} \frac{2 \pi}{H_{z \bar z}}Tr[...](z,\bar z)
      = \frac{1}{N_2} \int \frac {d^2t}{4} \frac{2 \pi}{H_{t \bar t}}Tr[...](t,\bar t)
\eea      
The contribution to $S_{YM}$ in the $ASD$ sector
maintains the same form because of its invariance
under conformal rescaling of the metric.
For the $SD$ part there might be additional
terms due to the fact that a holomorphic twist of the Higgs field in the $ASD$
sector is anti-holomorphic in the $SD$ one. 
The holographic quantum effective action reads (we do not display explicitly
the dependence on $\lambda$):
\bea
\Gamma_{q}= \Gamma |_{A_y=0}|_{FD} -\sum_p \sum_{i \ne j}logDet(ad\mu^+_p)|_{\mu_p^-=0}
\eea  
The first term, $\Gamma|_{A_y=0}|_{FD}=((\frac{N}{4g^2}S_{YM}
+\frac{1}{2}logDet'(-\Delta_A \delta_{ \mu \nu}
+ D_{\mu} D_{\nu} +i ad_{F^-_{ \mu \nu}})
-logDet'\frac{D\mu }{D\mu'})|_{A_y=0})|_{FD}$, is the "classical" holographic action 
associated to the reduced non-commutative theory
on the fundamental domain in the axial gauge.
From a computational point of view it is more convenient to calculate
the functional determinants in terms
of the original Hitchin system before the conformal map to the fundamental domain.
The two expressions for $\Gamma$ should coincide up to the conformal anomaly.
Thus $\Gamma |_{A_y=0}|_{FD}=
\Gamma |_{A_y=0}|_{Sphere}+Conformal Anomaly$.
The structure of $\Gamma |_{A_y=0}|_{Sphere}$ will be elucidated below.
The second term is due to the additional
gauge fixing to a Borel sub-algebra, $\mu^{-}_{p}=0$, and it is the $d=4$ analogue of the
Vandermonde determinant of the eigenvalues, to which it reduces in the unitary limit
$ \lambda \rightarrow 0$. The label $^-$ for $\mu_{p}$ means here lower
triangular part, excluding the diagonal, while the label $^+$ for $\mu_{p}$
means here upper triangular part including the diagonal.
The conformal anomaly is the contribution due to the change of metric implicit
in mapping conformally the region encircled by the punctured Wilson loop to
the cuspidal fundamental domain. It can be obtained from the exact
beta function.
The holographic loop equation on the fundamental domain reads (we do not
display explicitly the dependence on $\lambda$):
\bea
0=\tau(\frac{\delta \Gamma_{q}}{\delta \mu^{+(t)}_{p}}\Psi(x_p,x_p;B)) 
\eea
that is implied by the critical equation:
\bea
\frac{\delta \Gamma_{q}}{\delta \mu^{+(t)}_{p}}=0
\eea
which we refer to as the master equation. We should notice that, in
analogy with the two dimensional case, corrections due to the external region
and to the periodicity of the eigenvalues should be included.
We write now the "classical" holographic action, $\Gamma$, in terms of
functional determinants.
For the purpose of computing the first coefficient of the beta function it is
considerably more convenient to calculate $\Gamma$ in a Lorentz gauge rather
than $\Gamma_q$ in the axial gauge, that is needed to realize the reduction to a Borel
sub-algebra and, in the unitary limit, to a diagonal one.
We expect that the divergences of $\Gamma$ and $\Gamma_q$ coincide, since
$\Gamma_q$ is obtained from $\Gamma$ just by a conformal map and a suitable
gauge fixing. Since however the conformal map is singular, subtleties might be
involved and thus we would like to have a direct understanding
of $\Gamma_q$. Yet, we will not address this problem in this paper.
Part of the computation of $\Gamma$ in a Lorentz gauge appeared already in 
\cite{MB}. However we have computed here more explicitly than in \cite{MB}
the Jacobian of the change of variables to the holomorphic gauge.
This Jacobian turns out to be essential to reproduce
the correct value of $\beta_0$ in the limit
$\lambda \rightarrow 0$, which is the one relevant to the physical case
of Wilson loops for the unitary connection $A$. In this respect
the computation of the beta function that we present here is more
relevant to the physics than the one in \cite{MB}, where 
the Jacobian to the holomorphic gauge was not computed and
$SD$ of the master field implicitly assumed in the sector of $ASD$ type.
In this case it was found in \cite{MB} that the one-loop perturbative
beta function was accounted completely by the localisation determinant
(see next section).
In the Feynman gauge $\Gamma$ is the sum of the
classical (reduced, non-commutative) Yang-Mills action plus the logarithm of a 
number of determinants whose origin is as follows.
There is the determinant that arises from the localisation
integral on the microcanonical ensemble.
The localisation determinant is defined formally as:
\bea
\int DA_{\mu} \delta(F^-_{ \mu \nu}- \mu ^-_{ \mu \nu})= Det'^{-1}(P^- d_A \wedge)
\nonumber \\ 
=Det'^{-\frac{1}{2}} ((P^- d_A \wedge)^*(P^- d_A \wedge)) \nonumber \\
=Det'^{-\frac{1}{2}}(-\Delta_A \delta_{ \mu \nu} + D_{\mu} D_{\nu} +i ad_{F^-_{ \mu \nu}} )
\eea
where $ P^- $ is the projector onto the anti-selfdual part of the curvature.
The $ ' $ suffix requires projecting away from the determinants the zero modes due to gauge invariance, since
gauge fixing is not implied in the loop equation, though it may be understood
if we like to. \\
The careful reader may have noticed the unusual spin term, 
$i ad_{F^-_{ \mu \nu}}$, as opposed to $2i ad_{F_{ \mu \nu}}$ which arises
in the perturbative effective action (see next section).
The occurrence of $F^-_{ \mu \nu}$ is due to the projector $P^-$ in Eq.(54).
The coefficient of $F^-_{ \mu \nu}$ is one half of the one
of $F_{ \mu \nu}$
since the delta functional in Eq.(54)
is approximated by a Gaussian whose quadratic form is the second order
expansion of
$ (F^-_{ \mu \nu}- \mu ^-_{ \mu \nu})^2 $ around $\mu ^-_{ \mu \nu}$,
while the usual background field
method involves expanding the quadratic form associated to $F^2_{ \mu \nu}$
around the background $\mu _{ \mu \nu}$ (see next section).
After inserting the gauge fixing condition and the corresponding Faddeev-Popov
determinant the localisation determinant can be defined non-formally in the Feynman gauge as:
\bea
Det'^{-\frac{1}{2}}(-\Delta_A \delta_{ \mu \nu} + D_{\mu} D_{\nu} +i ad_{F^-_{ \mu \nu}} )
= \nonumber \\
= \lim_{\epsilon \rightarrow 0}
 \int Dc DA_{\mu} \exp(-\frac{1}{2 \epsilon}\int d^4x Tr(c^2)) \times \nonumber \\
\times \exp(-\frac{1}{4 \epsilon} \sum_{\mu \ne \nu}\int d^4x Tr((F^-_{ \mu \nu}-\mu ^-_{ \mu \nu})^2))) \delta(D_{ \mu} \delta A_{\mu}-c) \Delta_{FP}
\eea
The result is thus:
\bea
Det'^{-\frac{1}{2}}(-\Delta_A \delta_{ \mu \nu} + D_{\mu} D_{\nu} +i ad_{F^-_{ \mu \nu}} )=
 Det^{-\frac{1}{2}}(-\Delta_A \delta_{ \mu \nu}+i ad_{F^-_{\mu
\nu}}) \Delta_{FP}
\eea
where $ F^-_{ \mu \nu}$ is the anti-selfdual part of the field strength, given by:
\bea
F^-_{ \mu \nu}=F_{\mu \nu}-F^*_{ \mu \nu}
\eea
with:
\bea
F^*_{ \mu \nu}=\frac{1}{2} \epsilon_{ \mu \nu \alpha \beta} F_{ \alpha \beta}
\eea
In addition the determinant that arises as the Jacobian of the change of
variables to the holomorphic gauge contributes to the "classical" holographic action:
\bea  
-logDet'\frac{D\mu }{D\mu'}
\eea
which can be computed more explicitly as:
\bea
logDet(1+[G^{-1}\frac{\delta G}{\delta \mu},\mu])
\eea
where the following equation has been used:
\bea
G^{-1} \bar{\partial} G=-i \bar b
\eea
The "classical" holographic action, $\Gamma$, reads:
\bea
\Gamma= \frac{N}{4 g^2} S_{YM}{+\frac{1}{2}}logDet(-\Delta_A \delta_{\mu \nu}+i ad_{F^-_{\mu
\nu}})-log\Delta_{FP}-logDet \frac{D \mu }{D \mu'}
\eea
where $S_{YM}$ is the Yang-Mills action of the reduced non-commutative
theory and we assume a corresponding reduction in the other terms where
necessary. The normalisation of $\Gamma$ given by the reduced theory is
needed to be compatible with the two dimensional normalisation of the
Vandermonde-like determinant in Eq.(51). 
Notice that the last term in Eq.(62) vanishes if $\mu$
is reduced by gauge fixing to an Abelian or to a Borel sub-algebra,
since the commutator in Eq.(60) involves also the colour indices
and the commutator of elements in the Borel sub-algebra is nilpotent.

\section{Beta function: the first coefficient}

We compute in this section the first coefficient of the beta function as
it follows from the "classical" holographic effective action $\Gamma$.
In this section for simplicity we evaluate $\Gamma$ in the un-reduced theory,
rather than in the reduced one. The reduction is irrelevant as far as
$\beta_0$ is concerned, since its computation does not involve the second term
in Eq.(51). A partial computation involving only the part of $\Gamma$ composed
by the classical Yang-Mills action and the localisation determinant was already performed in
\cite {MB}. Following \cite{MB}, it is convenient to perform the computation in 
an indirect way, by means of a term by term comparison with the usual one-loop
perturbative contribution to the effective action.
For this purpose let us recall the structure of one-loop perturbative
corrections to the classical action in the Feynman gauge: 
\bea
\int Dc DA_{\mu} \exp(-\frac{N}{2 g^2} \int d^4 x Tr(c^2)) \exp(-\frac{N}{4 g^2} S_{YM}) \delta(D_{\mu} \delta A_{\mu}-c) \Delta_{FP}= \nonumber \\
=\exp(-\frac{N}{4 g^2}S_{YM})Det^{-\frac{1}{2}}(-\Delta_A \delta_{\mu \nu}+i 2 ad_{F_{\mu \nu}}) Det(-\Delta_A) 
\eea
where we have inserted in the functional integral the gauge-fixing condition and the corresponding
Faddeev-Popov determinant and, by an abuse of notation, we have denoted with $A$
the classical background field in the right hand side of Eq.(63). It follows that
the perturbative one-loop effective action in the Feynman gauge is given by:
\bea
\Gamma_{one-loop}= \frac {N}{4 g^2}S_{YM}+\frac {1}{2} 
logDet(-\Delta_A \delta_{\mu \nu}+i2ad_{F_{\mu
\nu}})- logDet(-\Delta_A )
\eea
The perturbative computation of the one-loop beta function \cite{AF,AF1} is the result of
two contributions that are independent within logarithmic accuracy \cite{Pol3}.
The orbital contribution gives rise to diamagnetism and to a positive 
term in the beta function:
\bea
-log(Det^{-\frac{1}{2}}(-\Delta_A \delta_{\mu \nu}) Det(-\Delta_A))=
log Det(-\Delta_A)= \nonumber \\
= \frac{1}{12} \frac{N}{(2 \pi)^2}log(\frac {
\Lambda}{\mu}) \frac{1}{2} \sum_{\mu \ne \nu} \int d^4x Tr(F_{\mu \nu})^2
\eea
where it should be noticed the cancellation of two of the four polarisations between
the first factor and the Faddev-Popov determinant.
The spin contribution gives rise to paramagnetism and to an
overwhelming negative term in the beta function \cite{Pol3}:
\bea
\frac{1}{4} \sum_{\mu \ne \nu} Tr(i2 ad_{F_{\mu \nu}} (-\Delta_A )^{-1}
i2 ad_{F_{\mu \nu}} (-\Delta_A )^{-1})= \nonumber \\
= 2 Tr( i ad(\mu^0) (-\Delta_A )^{-1} i ad(\mu^0) (-\Delta_A )^{-1})= \nonumber \\
= - \frac{12}{12} \frac{N}{(2 \pi)^2} log(\frac {\Lambda}{\mu})
 \frac{1}{2} \sum_{\mu \ne\nu}\int d^4x Tr(F_{\mu \nu})^2
\eea
where for later convenience we have expressed the spin contribution in the $ \lambda \rightarrow 0 $
limit in terms of the
field $\mu^0$.
Hence the complete result for the divergent part of $\Gamma_{one-loop}$ is:
\bea
(\frac {N}{2 g^2}- \frac{11}{12} \frac{N}{(2 \pi)^2}log(\frac {\Lambda}{\mu})) 
\frac {1}{2} \sum_{\mu \ne \nu} \int d^4x Tr(F_{\mu \nu})^2
\eea
from which it follows that:
\bea
\beta_0= \frac {11}{12} \frac {1}{(2 \pi)^2}
\eea
From Eq.(62) it can be easily read that the orbital contribution in 
$\Gamma_{one-loop}$ and $\Gamma$ coincide.
On the contrary, the spin contribution in $\Gamma$
involves only the anti-selfdual part of the curvature instead of
the curvature itself.
Hence the spin contribution
from the localisation determinant in the unitary limit is only one half of the spin
contribution in perturbation theory. Thus the orbital and spin contributions
of the localisation determinant sum up to: 
\bea
(\frac {N}{2g^2}-(\frac {1}{2}-\frac {1}{12})\frac{N}{(2 \pi)^2} log(\frac {\Lambda}{\mu}))
Tr(\mu^{02})
\eea
In fact, remarkably, the missing other one half in the coefficient of the beta function
is furnished by the Jacobian to the holomorphic gauge in
the unitary limit, which to second order in $ \mu^0 $ contributes to $\Gamma$:
\bea
logDet(1-i [\bar \partial^{-1} \frac{\delta \bar b}{\delta \mu}|_{\mu=\mu^0}, \mu^0])
\eea
where the commutator involves space-time indices and colour as well.
From the $ASD$ localisation it follows to lowest order in $\mu$:
\bea
\frac{\delta \bar b}{\delta \mu}=-i \bar \partial (- \Delta)^{-1}
+\bar \partial_u (-\Delta)^{-1}
\eea
The second term in the preceding equation does not contribute
to second order because of rotational and parity invariance 
in the $(u,\bar u)$ plane.
Eq.(70) contains traces of powers of a commutator, hence a formal
evaluation would hardly lead to any logarithmic divergence.
For example the contribution which we are interested in is the trace 
of a commutator squared:
\bea
-\frac{1}{2} Tr([(-\Delta)^{-1},ad(\mu^0)]^2)
\eea
where now the commutator involves only space-time indices.
Yet, one term is zero in dimensional regularization, being a tadpole:
\bea
Tr((-\Delta)^{-2},ad(\mu^0)^2)
\eea
while the other term:
\bea
- Tr((-\Delta)^{-1}ad(\mu^0)(-\Delta)^{-1}ad(\mu^0))
\eea
furnishes the needed contribution with the correct
sign and coefficient to combine with the result from localisation exactly into
the perturbative $\beta_0$.
This seems to be a non-trivial check of our chain of changes of variable.

\section{ Beta function: the contribution of the rescaling anomaly
to the second coefficient}

It is an interesting question as to whether
$\Gamma$ reproduces also the second coefficient of the
beta function. We do not have a definite answer at the moment.
A contribution to $\beta_1$, if it exists, can only come 
from higher order operators, starting with $ Tr({\mu}^4) $,
that would carry the correct power of $g$. It is then clear that, to relate
$ Tr({\mu}^4) $ to $ Tr({\mu}^2) $, Eq.(25) should be employed.
A shortcut may perhaps consist in choosing a $\mu$ that reproduces to lowest
order the gauge propagator as in the two dimensional case.
A most natural thing would be to compare our holographic
effective action with the Wilsonian effective
action of $QCD$, based on the exact renormalization
group \cite{Polc, Gall}, if we knew it. In principle an ansatz for the exact Wilsonian effective action
should include all higher order operators as it turns out in our holographic
case. Yet the simplest ansatz that includes only the lowest order operator
$ Tr (F_{\mu \nu}^2) $ was already considered in \cite{RW}, where it was found,
with this ansatz, the following beta function:
\bea
\frac{\partial g}{\partial log \Lambda}=
-\frac{g^3}{16 \pi^2} \frac{11}{3}(1-\frac{g^2}{16 \pi^2} \frac{7}{2})^{-1}
\eea
We would like to give an explanation of this result in the light of our
holographic effective action.
In the case that the Wilsonian effective action was truncated to the lowest order
operator, we should compare the result in \cite{RW} with ours without including the
higher order operators. In this case our holographic effective action is only
and exactly one loop as in the super-symmetric case. At first sight we have a discrepancy,
since our result is only one loop while \cite{RW} exhibits a NSVZ-like beta function
\cite{Shiff}.
However, following the analogy with the super-symmetric case,
we should remind that in \cite{RW} it has been computed
the canonically normalised Wilsonian effective action, while
we have computed the effective action with the Wilsonian non-canonical
normalisation.
We can pass from one to the other one by means of a Jacobian that takes into
account the rescaling anomaly, following \cite{HM}.
This Jacobian has been computed in the $ {\cal N}=1$
super-symmetric case and in the super-symmetric theory it
accounts 
for the link between the holomorphic beta function, that is only one loop, 
and the beta function for the canonical coupling.
Oddly the analogous computation does not seem to have been ever performed for pure $QCD$.
We fill here this gap. In fact our computation mimics the one
in the super-symmetric case, that is therefore recalled below.    
In the gauge theory with $ {\cal N}= 1 $ super-symmetry, without matter multiplets,
$log J$, the logarithm of the Jacobian for the rescaling anomaly, receives a
contribution from the gluons, $V$, from the gluoinos, $\psi, \bar \psi$, 
form the auxiliary scalar field, $D$,
and from the scalar field that implements the gauge-fixing condition
in the Feynman gauge, $c$:
\bea
log J=log g \lim_{M \rightarrow \infty}( Tr_V(\exp(-\frac{1}{M^2}
(-\Delta_A \delta_{\mu \nu}+i 2 ad_{ F_{\mu \nu}})) \nonumber \\
- Tr_{\psi}(\exp(-\frac{1}{M^2}
\gamma_{\mu} D_{\mu}))
- Tr_{\bar \psi}(\exp(-\frac{1}{M^2}
(\gamma_{\mu} D_{\mu})) \nonumber \\
+ Tr_{D}(\exp(-\frac{1}{M^2}
(-\Delta_A))
- Tr_{c}(\exp(-\frac{1}{M^2}
(-\Delta_A)))
\eea
Quite obviously in the pure gauge case we get:
\bea
log J=log g \lim_{M \rightarrow \infty}( Tr_V(\exp(-\frac{1}{M^2}
(-\Delta_A \delta_{\mu \nu}+i 2 ad_{ F_{\mu \nu}})) \nonumber \\
- Tr_{c}(\exp(-\frac{1}{M^2}
(-\Delta_A)))
\eea
We have computed that in this case:
\bea
log J
=log g \beta_J \frac{1}{2}\sum_{\mu \ne \nu}\int Tr(F_{\mu \nu}^2) \nonumber \\
=log g \frac{1}{16 \pi^2} \frac{7}{2} \frac{1}{2}\sum_{\mu \ne \nu}\int Tr(F_{\mu \nu}^2)
\eea
Now, since the canonical coupling, $g_c$, and the non-canonical Wilsonian coupling, $g_W$,
are related by the relation:
\bea
\frac{1}{2g_c^2}=\frac{1}{2g_W^2}- \beta_J log g_c
\eea
and the Wilsonian coupling is only one loop within our accuracy,
we get for the canonical beta function, within our accuracy:
\bea
\frac{\partial g_c}{\partial log \Lambda}=
-{g_c^3} \beta_0 (1-g_c^2 \beta_J)^{-1}
\eea
with:
\bea
\beta_0=\frac{1}{16 \pi^2} \frac{11}{3} \nonumber \\ 
\beta_J=\frac{1}{16 \pi^2} \frac{7}{2} 
\eea
in perfect agreement with the result in \cite{RW}.
From Eq.(80)-(81) we can read that the contribution of the rescaling anomaly to $ \beta_1$ in the pure 
gauge theory is:
\bea
\beta_{1J}=\beta_0 \beta_J= \frac{1}{(16 \pi^2)^2} \frac{77}{6}
\eea
where our convention is that positive $\beta_0, \beta_1$ give origin to a negative beta
function. Incidentally this seems to answer a question raised by Shifman
\cite{Shiff}
about the operator versus canonical conformal anomaly in the pure gauge theory
(within our accuracy, since in principle there could be contributions of higher
order operators also in the Jacobian $J$).
This shows that the missing term that must come from higher order
operators to get
agreement with the two-loop perturbative result:
\bea
\beta_1=\frac{1}{(16 \pi^2)^2} \frac{68}{6}
\eea
is:
\bea
\beta_{1W}=- \frac{1}{(16 \pi^2)^2} \frac{9}{6}
\eea
We leave for the future the evaluation of higher order operators
in our holographic effective action.

\section{Miscellanea}

In this section we collect some comments about analogies with the existing literature.
In recent years there has been much progress in understanding the non-perturbative
physics of super-symmetric gauge theories based on such concepts as effective action,
holomorphy, holography and integrable systems.
One of the first non-trivial results was the exact NSVZ (canonical) beta function
\cite {Shiff} in $ {\cal N}=1$ super-symmetric theories, that is related to the fact that
the Wilsonian (holomorphic) coupling constant gets only one-loop divergences because of the
holomorphic
properties of the ${ \cal N}=1$ $SUSY$ action written in terms of super-fields.
As we have just seen this distinction plays a role in this paper, if we want to relate
our Wilsonian holographic effective action to the perturbative canonical one
by a computation analogous to the one in \cite{HM}.
Some time ago a holographic correspondence was found in \cite{Mal}, between a 
boundary $ {\cal N}=4$
four dimensional super-symmetric gauge theory and a bulk five dimensional super-gravity
(super-string) theory.
New techniques based on integrable systems and holomorphic matrix models
have been developed to find the low energy effective action of
$ { \cal N} =2$ and $ {\cal N}=1$ super-symmetric gauge theories
in \cite{SW} and \cite{Vafa} respectively.
The extension of these techniques to non-supersymmetric gauge theories in four
dimensions such as $QCD_4$ encounters considerable difficulties.
Though at technical level the construction in this paper of the holographic effective action
in large-$N$ $QCD_4$ for planar self-avoiding loops seems peculiar, at least for the moment,
to the ${\cal N} =0$ theory,
we would like to interpret the existence of such effective action
in the light of some of the concepts that played such an important role
in the super-symmetric case.
We have already stressed that the construction of the effective action
from the loop equation is a case of holography, i.e. of a correspondence between
a boundary theory supported on loops, defined by the loop equation, and a bulk theory
defined by the effective action.
The key point for the existence of this holographic correspondence is holomorphy, that
enters here in changing variables in the loop equation in such a way that the quantum 
contribution to the loop equation in these new variables involves the contour integral
of the Cauchy kernel
and thus can be reduced to the computation of a regularized residue.
In this respect the Cauchy theorem can be considered the
oldest and simplest case of holography. 
Yet, there are two other points of contact with the holography in the sense of \cite {Mal}.
The first one is that to make the quantum residue vanishing is necessary to map
the local bulk degrees of freedom of the theory into the boundary cusps.
Thus we have a correspondence between the degrees of freedom in the bulk
and at the boundary, that is the distinguishing feature of holography
in modern sense. In addition this is achieved by means of a conformal mapping
that induces a new metric on the fundamental domain in the upper half plane, in analogy with the $AdS$ theory \cite {Mal}.
Notice also that in our approach, in the four dimensional theory, there are logarithmic corrections
to such metric due to the conformal anomaly, as already suggested in \cite{Pol2}.
It is perhaps remarkable that these features arise directly from
first principles, i.e. directly from the loop equation. \\
The usual unitary monodromy
of the connection that enters the loop equation is embedded into a non-unitary family
of monodromies whose curvature is either of $ASD$ or of $SD$ type: this non-unitary family
plays the analogous role of 
the holomorphic chiral ring in $ {\cal N}=1$ super-symmetric gauge theories
\cite{CD}.
In fact the structure of the "classical" holographic action for large-$N$ $QCD_4$ 
itself has many analogies with the one
of $ {\cal N}=1$ super-symmetric gauge theories \cite{Vafa}.
This effective action is the sum of three terms.
The classical action, the Veneziano-Yankielowicz \cite{Ve} potential
(essentially the one-loop contribution) and the contribution of the Konishi anomaly, that occurs
as an anomaly in the conformal rescaling of the chiral super-fields in the functional integral.
It is interesting to observe that our "classical" holographic action has in fact
a very similar structure. The $ {\cal N}=1$ effective action is defined as a functional of the
glueball super-field, that is a scalar composite operator.
In our $ {\cal N}=0 $ case the "classical" holographic action is a functional of a composite operator
as well,
the self-dual and anti-selfdual (in the $Z_2$ orbifold version) components of the curvature tensor, that play
the role of the chiral curvature super-fields $W$ and $ \bar W$ in the super-symmetric case.
There is even a hyper-Kahler structure, often associated to super-symmetry,
that is inherited by the moduli fields of the Hitchin systems determined by
these $ASD$ or $SD$ fields.
Also the "classical" holographic action consists of three terms: the classical action,
the localisation determinant (that is only one loop) and the Jacobian of the change of variables to 
the holomorphic gauge. This third term is the analogue of the Konishi anomaly
and it is indeed a Jacobian under field-dependent gauge transformations living 
in the complexification of the gauge group.
The occurrence of Hitchin systems in the holographic effective action
can be considered the analogue of the occurrence of Hitchin systems in the 
solution for the low energy effective action of the Coulomb branch of $ {\cal N}=2$
super-symmetric gauge theories \cite{SW}.
We would like also to recall the reader that the idea of using Hitchin
systems to get control over the large-$N$ limit of $QCD$ arises by an energy-entropy
argument in \cite{MB1}, that is rooted in the idea of dominating the
large-$N$ limit by the saddle-point method. Hitchin systems, being completely integrable,
admit an "Abelianization"
map, not to be confused with the one of section 3, that maps parabolic $SU(N)$ Hitchin bundles into
$U(1)$ bundles over branched $N$ coverings. 
Thus the system gets Abelianized in such a way that the number of local moduli
passes from order of $N^2$ for the parabolic Hitchin system to order of $N$ for the Abelian
system on a $N$ covering. Hence for large-$N$ the entropy of the functional integration is re-absorbed
into the Jacobian of the change of variables of the Abelianization map and the usual saddle-point
argument for vector-like models applies.
In fact we could interpret the holographic map as an attempt to understand in a more precise way,
i.e. directly from the loop equation, how Hitchin systems help to get control over the
large-$N$ limit of $QCD$. 
Finally, the use that we have made of the conformal map to a cuspidal
fundamental domain resembles geometric engineering \cite{Vafa2} and in particular the fact
that only the local behaviour close to the singularity seems to count.
The use of cuspidal singularities resembles also the idea \cite{Pol2} of considering five dimensional
strings with the four dimensional boundary at a singular point of the metric in the
fifth coordinate, as required by the zig-zag symmetry. In turn the zig-zag symmetry plays a crucial role
in the vanishing of the quantum residue in the loop equation.

\section{Conclusions}

We here advocate a drastic change in our way of attempting a solution 
of the loop equation in the large-$N$ limit of $QCD$.
Rather than trying to solve this equation on the entire loop algebra, a highly
non-commutative problem, perhaps of non-hyperfinite nature,
we restrict ourselves to the commutative hyper-finite algebra generated by a 
fixed self-avoiding loop.
We look at this problem as a problem in holography, following the analogy
with the Cauchy theorem, that reconstructs from the values of a 
holomorphic function on a loop its values in the internal region.
We have constructed a holographic quantum effective action for the eigenvalues of
the curvature of a twisted local system, i.e. a centrally extended
Hitchin system. Hitchin systems occur in the functional integral by means of 
the localisation on a microcanonical ensemble labelled by the levels of the
$ASD$ ($SD$) curvature of the gauge connection.
We have obtained a holographic form of the loop equation by a change of variable from the
connection 
to the $ASD$ ($SD$) curvature in the holomorphic gauge. 
We have written the holographic loop equation for a special connection, whose
curvature is of $ASD$ ($SD$) type.
In the new integration variable, i.e. the $ASD$ ($SD$) curvature in the holomorphic gauge,
the quantum contribution to the loop equation
is reduced to the computation of a regularized residue, that turns out to be
loop independent for self-avoiding loops. To this form of the loop
equation we have associated a holographic "classical" effective action.
Finally, we have obtained the holographic quantum effective action
from the "classical" one by means of a conformal mapping of the
loop equation to a cuspidal fundamental domain on which the 
regularized residue for self-avoiding loops vanishes identically.
The conformal map makes possible a peculiar gauge choice
and a consequent reduction of the curvature to a Borel sub-algebra.
The physically interesting unitary Wilson loops are embedded into the
algebra generated by Wilson loops of $ASD$ ($SD$) type
via a limiting procedure that involves a twisting of the Higgs field of the
$ASD$ ($SD$) Hitchin system. \\
In this paper we have computed the first coefficient of the beta function
that arises from the holographic "classical" effective action for the connection  
in the unitary limit, finding exact agreement with the perturbative one-loop result. 
In addition we have computed that part of the second coefficient which, following
the analogy with the rescaling anomaly in $ {\cal N}=1$
super-symmetric gauge theory, arises from the rescaling in the holographic effective
action, in passing from the Wilsonian to the canonical coupling constant.
We found also in this case exact agreement with an independent result based on
continuum exact renormalization group.
The complete second coefficient of the beta function of the holographic
effective action is left as a problem for the future.

\end{document}